\documentclass{rmf-d}
\usepackage{nopageno,rmfbib,multicol,times,epsf,amsmath,amssymb,cite}
\usepackage[T1]{fontenc} 
\usepackage{caption2}
\usepackage{graphicx}
\usepackage{blindtext}
\usepackage{color}
\usepackage{hyperref}

\definecolor{UTOrange}{rgb}{1, 0.51, 0.0}

\usepackage{lineno}
\def\rmfcornisa{ \hfill\rmf\ {\bf 3} 040909 (2022) 1--4
\hfill September 2022}

\def\rmfcintilla{{\it Supl. Rev. Mex. Fis.} {\bf 3} 040909}
\clearpage \rmfcaptionstyle 
\begin{document}
\markboth{ANTONIO CARLOS OLIVEIRA DA SILVA}{ A \LaTeX template for the RMF, RMF-E, SRMF }

%
%
\title{Rivet and the analysis preservation in heavy-ion collisions experiments
\vspace{-6pt}}
\author{Antonio Carlos Oliveira da Silva (for the ALICE Collaboration)}
\address{University of Tennessee, Knoxville, 1408 Circle Drive, Knoxville TN 37996-1200 }
%
%
\maketitle
%
%
\recibido{3 July 2022}{15 September 2022
\vspace{-12pt}}
\begin{abstract}
\vspace{1em} 
%
%
The comparison of experimental data and theoretical predictions is important for our understanding of the mechanisms for interactions and particle production in hadron collisions, both at the Large Hadron Collider and at the Relativistic Heavy-Ion Collider experiments. Several tools were ideated to help with that. Rivet (Robust Independent Validation of Experiment and Theory) is a framework that facilitates the comparison between measurements from high-energy physics experiments and Monte Carlo event generators able to produce outputs using the HepMC package. Rivet contains a repository with analysis algorithms developed by experiments, providing analysis documentation and preservation.

The recent developments for the implementation of centrality and multiplicity classes in Rivet are presented in this contribution. 
\vspace{1em}
\end{abstract}
\keys{ \bf{\textit{
}} \vspace{-8pt}}
\begin{multicols}{2}

\section{Introduction}

Currently, the data and analysis preservation in high-energy physics experiments is becoming a common concern. Previous experiments and collaborations are losing the power of reproducing their measurements since the data are not properly kept in a accessible way. The old code, which contains crucial and detailed information
like detector acceptance, particle and event selections, and corrections, is no longer maintained and it is very difficult, if possible, to be run again. Comparisons of previous measurements with new models is, therefore, very challenging.

Robust Independent Validation of Experiment and Theory (Rivet)~\cite{buckley2010rivet} is a framework that aims to facilitate the comparison between data and Monte Carlo (MC) event generators.

\section{Rivet framework}

\begin{figure}[H]
 \includegraphics[width=\linewidth]
    {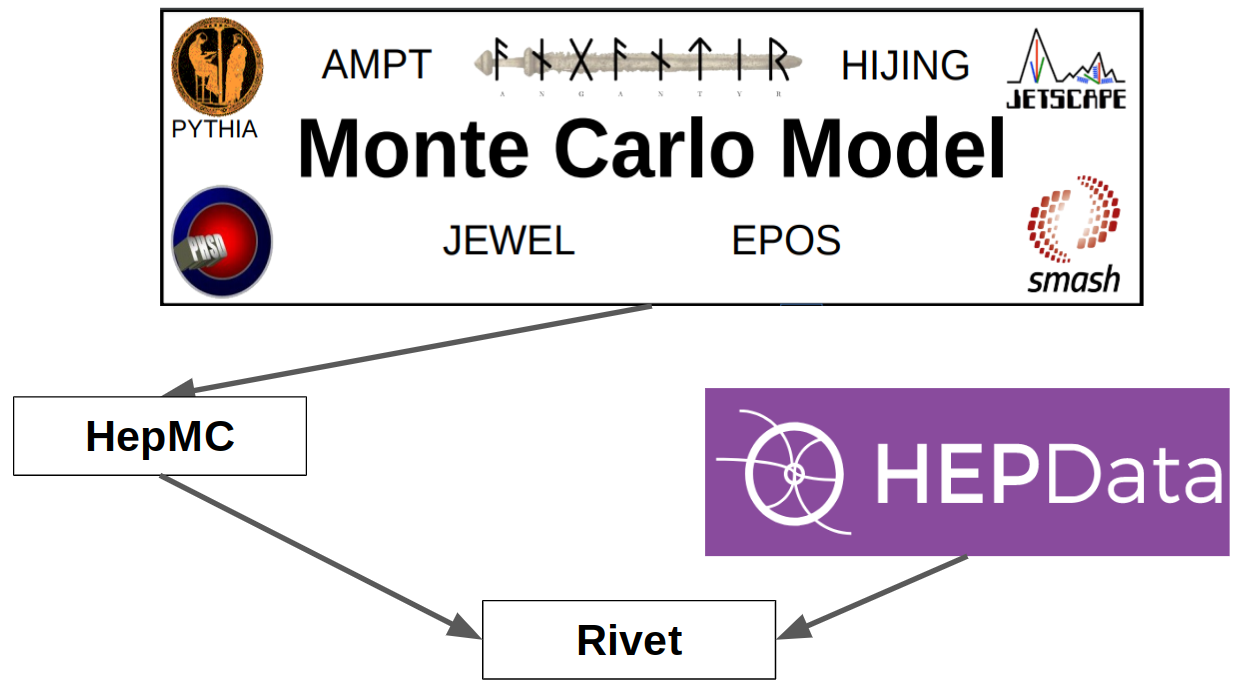}
 \caption{Schematic diagram showing how Rivet is integrated with HepMC and HepData.}
    \label{fig:RivetScheme}
\end{figure}

Rivet analyses are written in C++ and it currently contains more than 1000 analyses from several high-energy physics collaborations. The data, when available, are downloaded directly from HepData~\cite{Maguire_2017}. Any model that is incorporated in an event generator able to produce output that complies with HepMC framework~\cite{Dobbs:684090} can be used by Rivet for the comparison with data. The integration of Rivet with HepMC and HepData is pictured in the scheme presented in Fig. \ref{fig:RivetScheme}. 

The references for the event generators in Fig. \ref{fig:RivetScheme} can be found in\cite{Sjostrand:2007gs,AMPT,Bierlich:2018xfw,JETSCAPE,SMASH,EPOS,JEWEL,HIJING,PHSD}. Not all of them provide output using HepMC standards.

In principle, an article presenting a measurement should present enough information to make the measurement able to be reproduced by another experiment or theoretician interested in comparing the data with a model. However, some subtle details about detector acceptances, particle selections, trigger conditions, etc, could be missing or not clearly described. This can be the case even in internal notes in large collaborations.

Rivet aims at preserving all the analysis details. Furthermore, it should reproduce the methods used in a measurement as close as possible. One of the pillars of the Rivet philosophy is that we should treat Monte Carlo simulations in the same way as data.

In order to assure maximum fidelity to the original acceptances, selections and methods used in a measurement, whenever possible, the rivet analysis should be implemented by the collaboration that published the article containing the measurement. Currently, ALICE has an official internal procedure for Rivet analysis approvals.

\section{Centrality and multiplicity determination}

Measurements in heavy-ion collisions are commonly differential in centrality intervals to study different physical phenomena. Therefore, centrality determination is a crucial feature Rivet has to provide in order to reproduce measurements in heavy-ion collisions.

Centrality determination in ALICE is commonly provided by the V0 detector~\cite{collaboration_2013}, which consists of two arrays, V0-A and V0-C covering the pseudorapidity ranges 2.8 $<~\eta <~$ 5.1 and 3.7 $<~\eta <~$ 1.7 respectively.

\begin{figure}[H]
 \includegraphics[width=\linewidth]
    {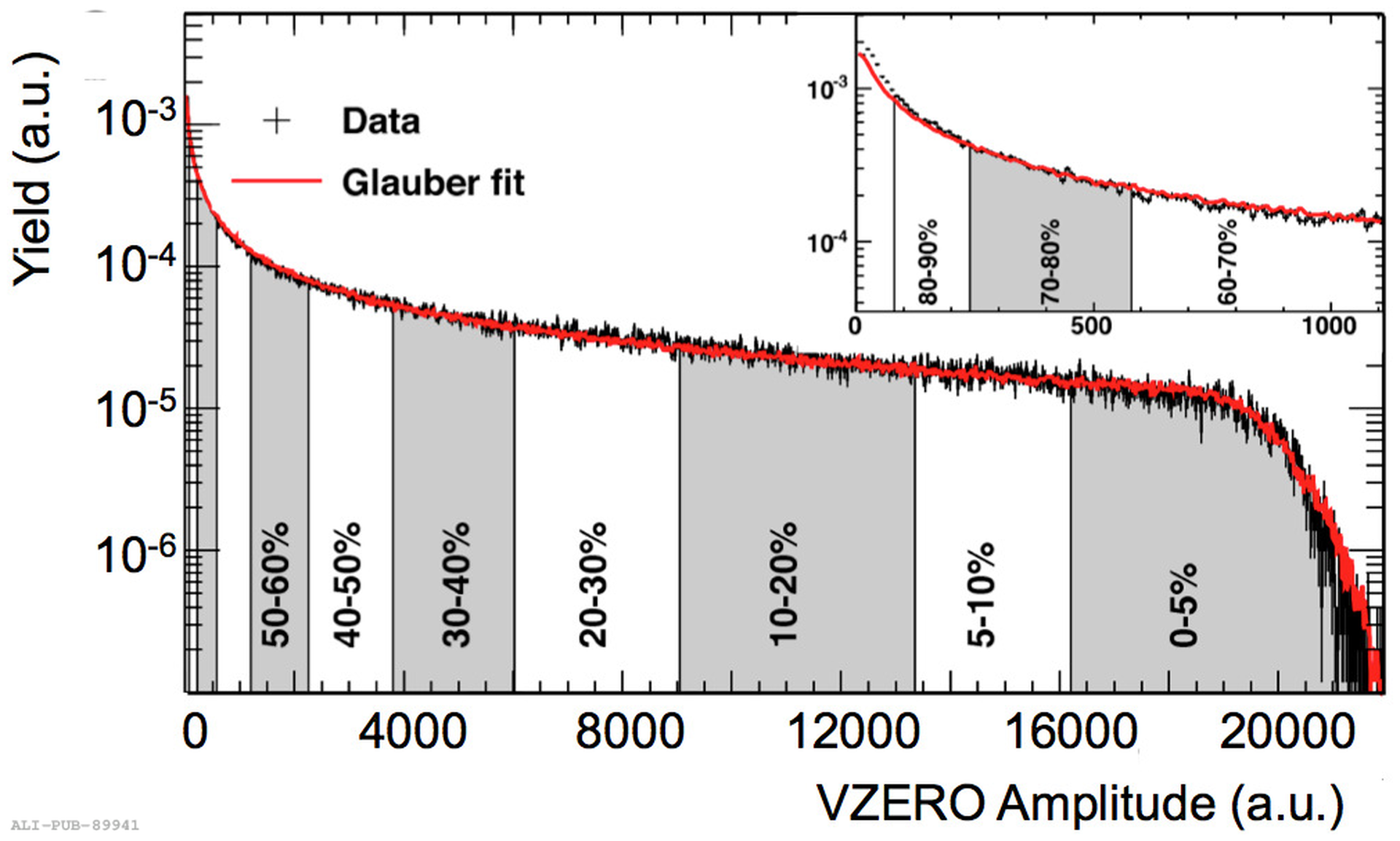}
 \caption{Distribution of the total amplitude in the V0 scintillators in black points. The data are fitted using a Negative Binomial Distribution (NBD) using parameters from the Glauber model.}
    \label{fig:CentDeterm}
\end{figure}

Figure \ref{fig:CentDeterm} presents the distribution of the total energy deposited in the V0 scintillators (amplitude). The most central collisions are associated with those event with highest V0 amplitude. The details of the centrality determination are described in~\cite{Abelev_2013}.

\begin{figure}[H]
 \includegraphics[width=\linewidth]
    {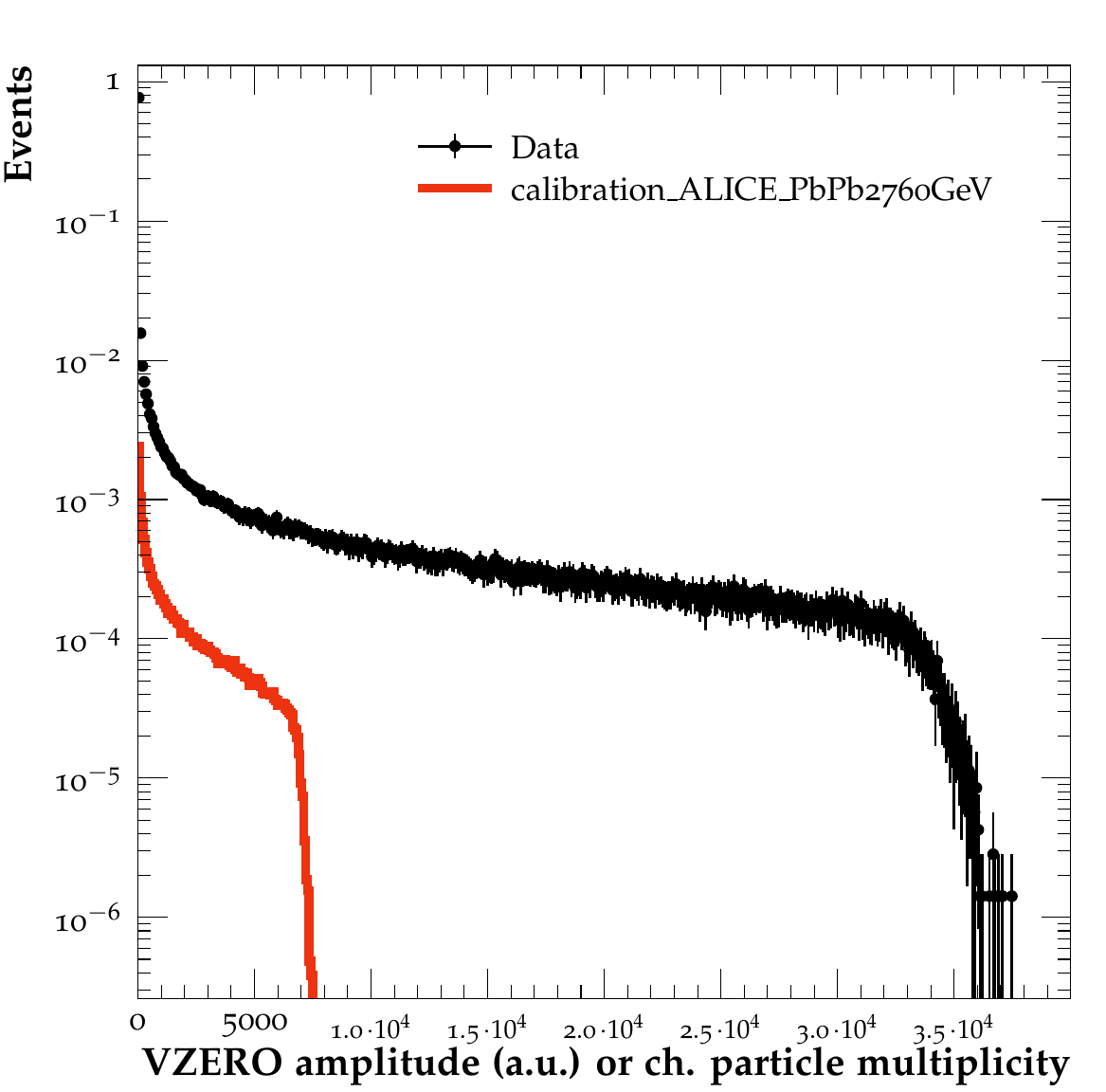}
 \caption{V0 amplitude in arbitrary units (black markers) measured by ALICE and the charged particle multiplicity in the V0 acceptance calculated with Rivet from Pb--Pb collisions at 2.76~TeV events simulated with PYTHIA~8 Angantyr (red line).}
    \label{fig:V0M}
\end{figure}

The centrality determination in Rivet uses the multiplicity of charged particles in the acceptance of the V0. Since the multiplicity of particles in heavy-ion collision events in Monte Carlo event generators is model dependent, it is necessary to create a calibration file that depends on collisions system, energy and event generator. Figure \ref{fig:V0M} shows the V0 amplitude distribution measured by ALICE and the multiplicity of charged particles in Pb--Pb collisions at $\sqrt{s_{\rm NN}} = 2.76$~TeV generated with PYTHIA~8 Angantyr~\cite{Bierlich:2018xfw}.

Rivet divides the charged particle multiplicity presented in figure \ref{fig:V0M} (red line) in centrality percentiles. So the most central events are associated to the highest multiplicity events. When running the Rivet analysis that requires centrality determination, this calibration has to be provided. A similar strategy is used for multiplicity determination in pp and p--Pb collisions. Currently, ALICE is developing the possibility to characterize the event using the self-normalized charged particle multiplicity distribution. This development is presented in sections \ref{SelfNormalized} and \ref{Results}.

\section{Self-normalized multiplicity}
\label{SelfNormalized}

\begin{figure}[H]
 \includegraphics[width=\linewidth]
    {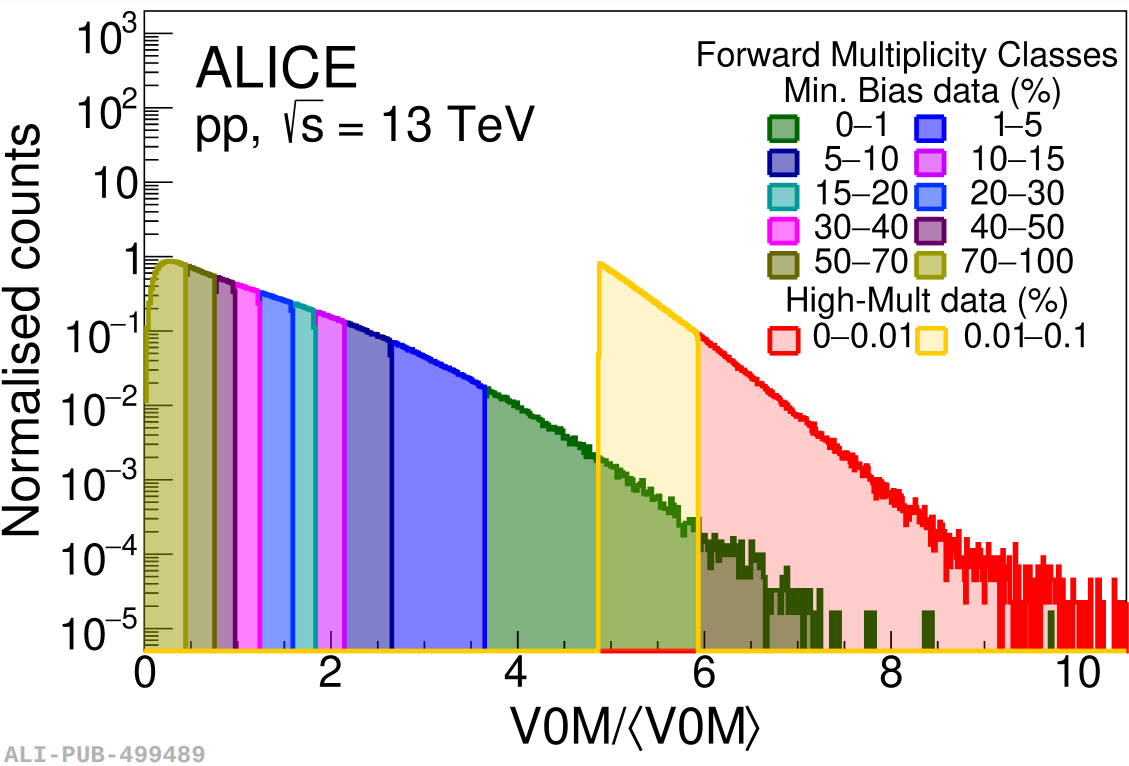}
 \caption{Distribution of the V0M amplitude scaled by its average value $\langle$V0M$\rangle$ used to determined forward-rapidity multiplicity classes in pp collisions at $\sqrt{s} = 13$~TeV.}
    \label{fig:SelfNormV0M}
\end{figure}

Forward-rapidity multiplicity classes can be defined in \mbox{ALICE} using the V0 detector. Figure \ref{fig:SelfNormV0M} shows the distribution of the V0M amplitude, which is proportional to the number of charged particles passing through the V0A and V0C detectors, scaled by its average value $\langle$V0M$\rangle$~\cite{SelfV0M2021}.

\begin{figure}[H]
 \includegraphics[width=\linewidth]
    {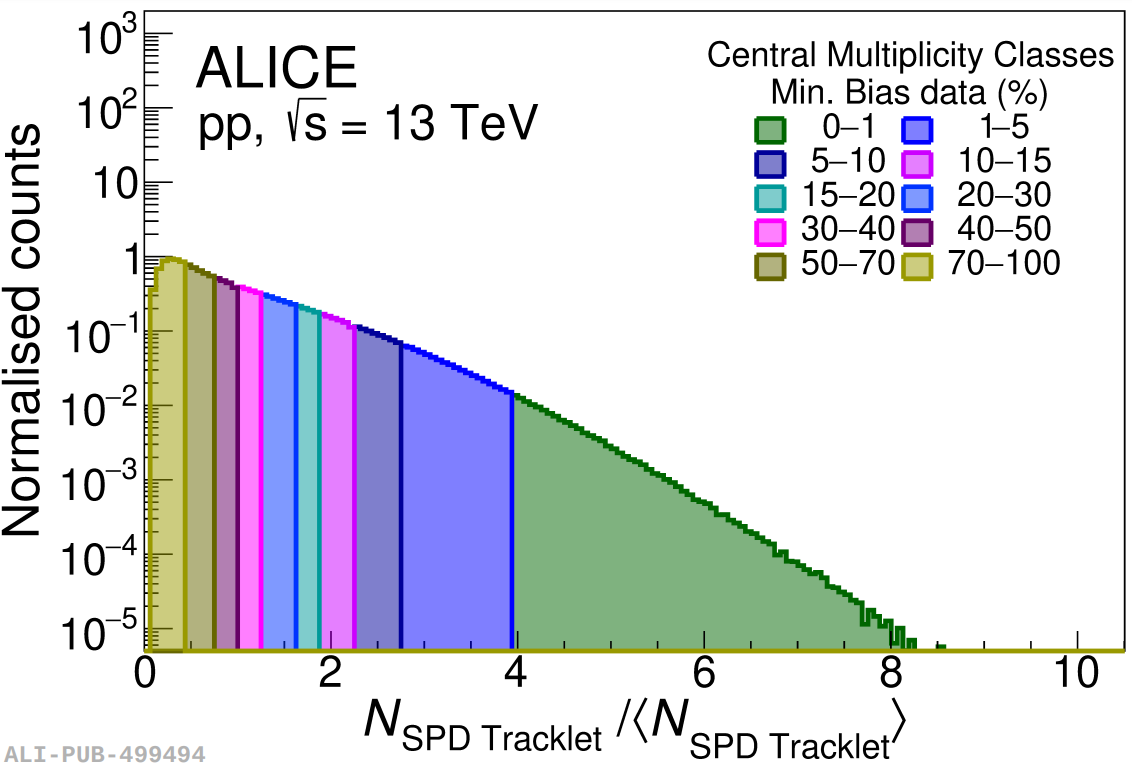}
 \caption{Distribution of the number of SPD tracklets scaled by its average value $\langle N_{\rm SPD~tracklets}\rangle$ used to determine midrapidity multiplicity classes in pp collisions at $\sqrt{s} = 13$~TeV.}
    \label{fig:SelfNormSPD}
\end{figure}

The Silicon Pixel Detector (SPD)~\cite{CERN-LHCC-99-012,Santoro_2009} is the closest detector to the interaction point in ALICE. The SPD provides mid-rapidity multiplicity classes determination using the reconstructed tracklets, which are track segments that connects hits in the two SPD layers pointing to the primary vertex. The self-normalized estimator is obtained with the distribution of SPD tracklets $N_{\rm SPD~tracklets}$ in -2 < $\eta$ < 2 scaled by the average of its value $\langle N_{\rm SPD~tracklets}\rangle$. The self-normalized SPD tracklets distribution is presented in Fig. \ref{fig:SelfNormSPD}.

\begin{figure}[H]
 \includegraphics[width=\linewidth]
    {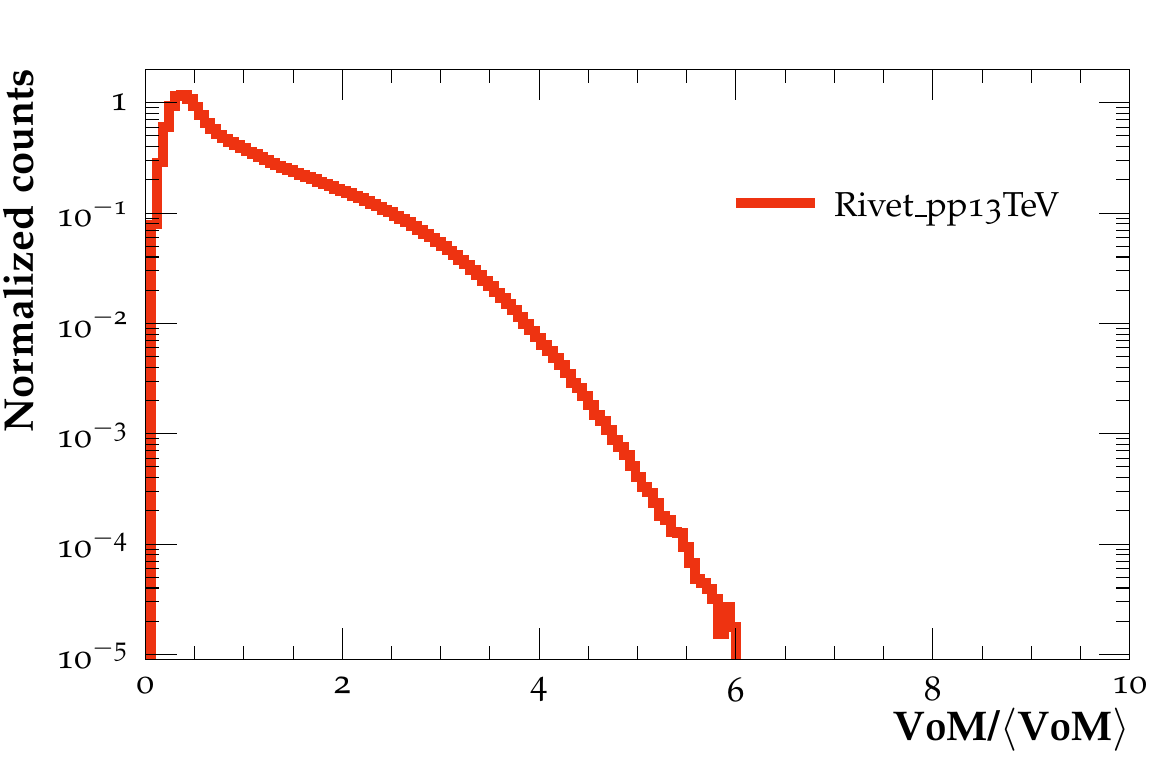}
 \caption{Self-normalized multiplicity distribution of charged particles in the acceptance of the V0 detector in pp collisions at $\sqrt{s} = 13$~TeV generated with PYTHIA~8 Monash~2013.}
    \label{fig:SelfNormV0MRivet}
\end{figure}

The event characterization using self-normalized multiplicity estimators in Rivet is under development in ALICE. Similar to what was discussed in the previous section, instead of using the V0M amplitude, Rivet uses the charged particle multiplicity in the V0 acceptance. In particular, the forward-rapidity self-normalized estimator uses the multiplicity of charged particles in the V0 acceptance scaled by its average value. Figure \ref{fig:SelfNormV0MRivet} shows the V0M/$\langle$V0M$\rangle$ distribution calculated with Rivet using pp collisions at 13 TeV generated with PYTHIA~8 Monash~2013 tune~\cite{Sjostrand:2007gs,Skands_2014}.

\begin{figure}[H]
 \includegraphics[width=\linewidth]
    {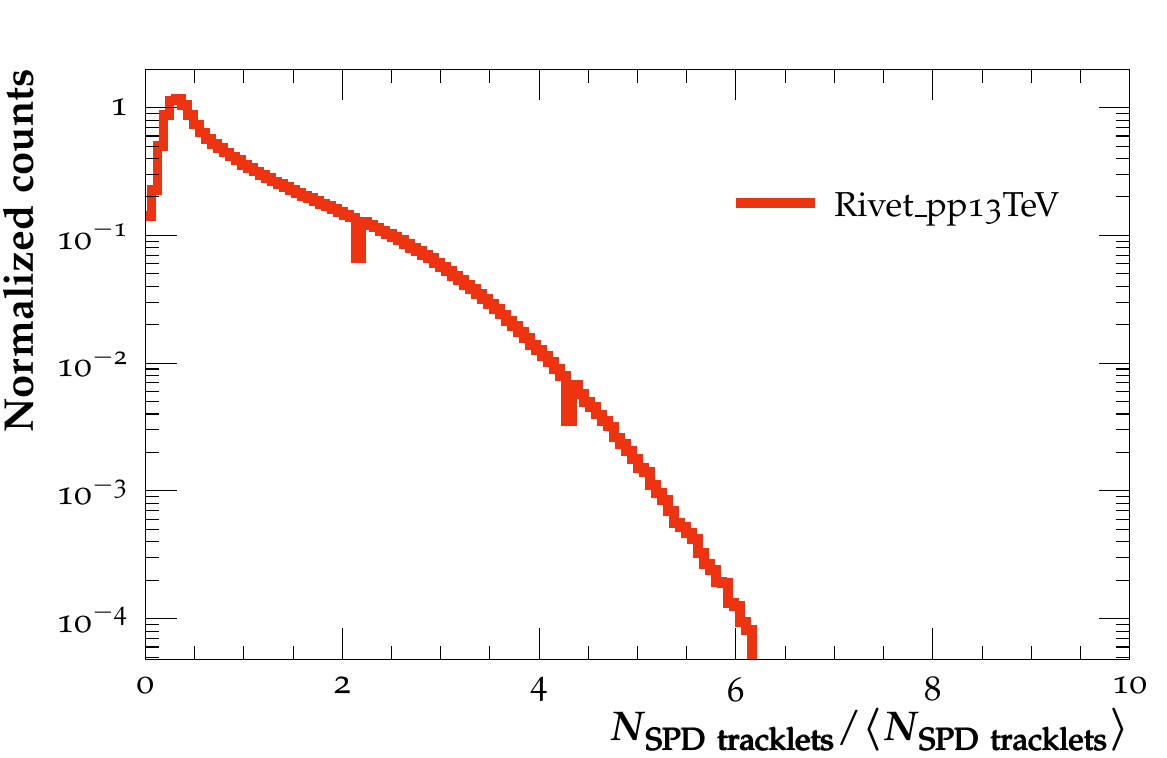}
 \caption{Self-normalized multiplicity distribution of charged particles in the acceptance of the SPD detector in pp collisions at $\sqrt{s} = 13$~TeV generated with PYTHIA~8 Monash~2013.}
    \label{fig:SelfNormSPDRivet}
\end{figure}

Similarly to what is done for the V0M, the self-normalized multiplicity estimator at mid-rapidity uses the number of charged particles in the acceptance of the SPD. Figure \ref{fig:SelfNormSPDRivet} shows the distribution of the charged particles in the SPD acceptance scaled by its average value in pp collisions at 13 TeV generated with PYTHIA~8 Monash~2013.

\section{Results using the self-normalized multiplicity estimators}
\label{Results}
The self-normalized multiplicity estimators framework in Rivet is currently work in progress and being tested using articles published by ALICE that use such estimators. The first measurement used to test the V0M/$\langle$V0M$\rangle$ estimator was the transverse momentum ($p_{\rm T}$) of jets in different multiplicity intervals in pp collisions at $\sqrt{s} = 13$~TeV~\cite{JetsInMult}. 

\begin{figure}[H]
 \includegraphics[width=\linewidth]
    {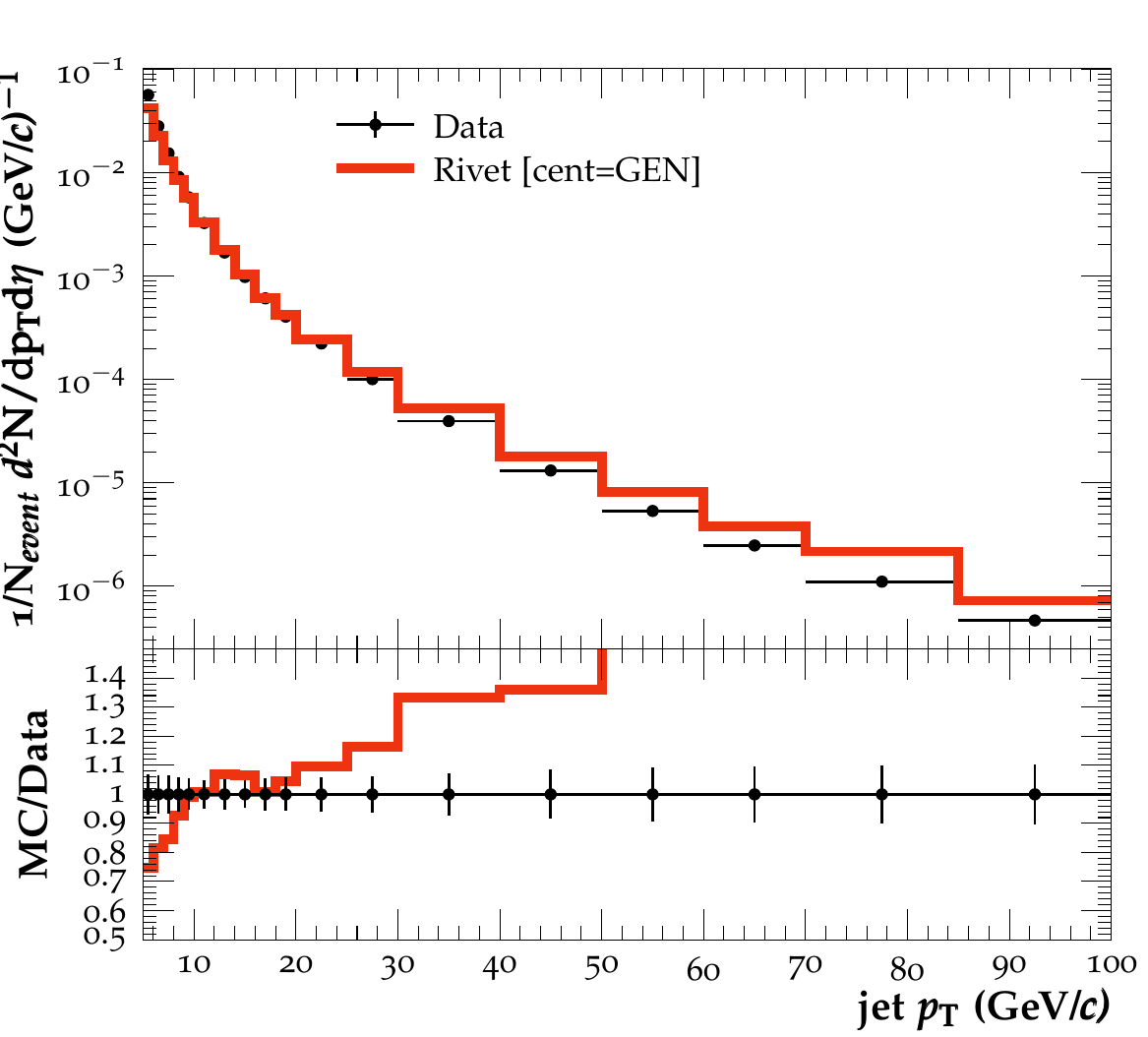}
 \caption{Charged-particle jet transverse momentum distribution in pp collisions at $\sqrt{s}=5.02$~TeV for the 0-1\% multiplicity class corresponding to the self-normalized V0M-based multiplicity estimator. Jets were reconstructed using jet resolution parameter $R$ = 0.2. Data (black markers) are compared with PYTHIA~8 Monash~2013 (red line).}
    \label{fig:JetsInMult}
\end{figure}

\begin{figure}[H]
 \includegraphics[width=\linewidth]
    {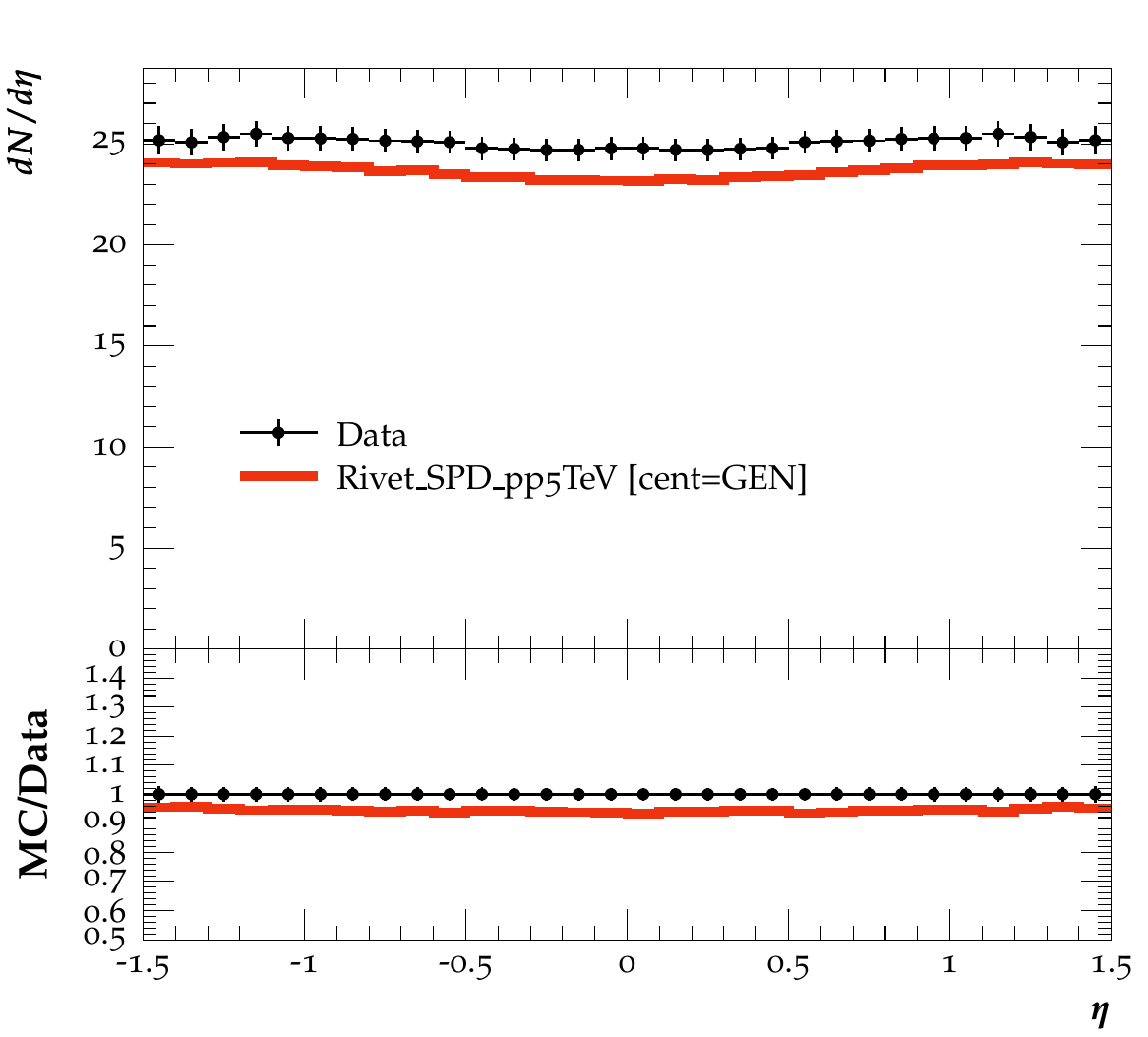}
 \caption{Charged particle pseudorapidity distribution in pp collisions at $\sqrt{s}=5.02$~TeV for the 0-1\% multiplicity class corresponding to the self-normalized SPD-based multiplicity estimator. Data (black markers) are compared with PYTHIA~8 Monash~2013 (red line).}
    \label{fig:EtaInMult}
\end{figure}

The transverse momentum of jets reconstructed with FastJet anti-$k_{\rm T}$ algorithm~\cite{Cacciari_2012} and resolution parameter $R$ = 0.2 in the multiplicity class 0-1\% in pp collisions at $\sqrt{s} = 13$~TeV is presented in figure \ref{fig:JetsInMult}. The measurement was compared with PYTHIA~8 Monash~2013 using Rivet and the self-normalized V0M multiplicity framework. The agreement of the model to data is a positive indication that the framework can reproduce the multiplicity determination method used by ALICE. Other multiplicity classes presented a similar performance.

The self-normalized estimator using the SPD was also tested using the measurements in~\cite{SelfV0M2021}. Figure \ref{fig:EtaInMult} presents the charged particle pseudorapidity distribution in pp collisions at $\sqrt{s} = 5.02$~TeV. The ALICE data are compared with PYTHIA~8 Monash~2013 using Rivet and the SPD self-normalized framework. The results fairly reproduce the comparisons to MC presented in the cited article.

\section{Summary}

Rivet is a valuable tool for analysis preservation and comparison of data to Monte Carlo event generators. The development of additional tools to facilitate the implementation of Rivet analyses is an important contribution to the framework and can be of benefit both for the experiment and the theory side. The self-normalized multiplicity estimators are providing consistent results between MC curves in Rivet and those provided by experiments. The final goal is to make this multiplicity framework available soon in the Rivet official framework.

\noindent

\end{multicols}

\medline
\begin{multicols}{2}
%
\nocite{}
\bibliographystyle{rmf-style}
\bibliography{Rivet}

%
%
%

%
%
\end{multicols}
\end{document}